\documentclass[aps,twocolumn,groupedaddress]{revtex4-2}

\usepackage{epsfig}
\usepackage{amsmath}
\usepackage{braket}
\usepackage{xspace}
\usepackage{graphicx}
\usepackage{lineno}

\begin{document}
\draft
%\twocolumn[\hsize\textwidth\columnwidth\hsize\csname@twocolumnfalse\endcsname

\title{Gate-voltage-driven quantum phase transition at  $0.7 (2e^2/h)$ in quantum point contacts}

\author{Jongbae Hong}
\affiliation{Asia Pacific Center for Theoretical Physics, Pohang, Gyeongbuk 37673, Korea}
\date{\today}

\begin{abstract}
We investigate a quantum phase transition (QPT) in quantum point contacts by analyzing the gate-voltage-dependent quasiparticle energy at the Fermi level at zero temperature. This energy is computed using the local density of states at the site of the localized spin, which is extracted from the replicated gate-voltage-dependent differential conductance shaped by entangled-state tunneling. The QPT occurs between symmetric (\(G \geq 0.7 G_0\)) and asymmetric (\(G < 0.7 G_0\)) Kondo coupling states, where \(G_0 = 2e^2/h\), and is driven by the migration of a localized spin in response to the side-gate voltage. The asymmetric state exhibits two distinct Kondo temperatures, while the symmetric state has only one. The existence of two Kondo temperatures in the \(G < 0.7 G_0\) regime accounts for both the anomalous gate-voltage dependence of the zero-bias anomaly width and the inability to define a Kondo temperature in the \(G < 0.7 G_0\) region.

\end{abstract}


\maketitle \narrowtext 

\section{Introduction}
The Kondo effect in nanoscale systems was first observed in quantum dots~\cite{Goldhaber,Goldhaber-PRL}, followed by quantum point contacts (QPCs)~\cite{Cronenwett}. 
In both systems, Kondo temperatures were determined through scaling analysis~\cite{Goldhaber-PRL,Cronenwett}, providing compelling evidence of Kondo physics.
However, several puzzling features remain unresolved. 
In particular, for QPCs, the scaling analysis breaks down below $0.7 G_0$ in conductance—where $G_0=2e^2/h$, with $e$ the electron charge and $h$ Planck’s constant—making it impossible to determine the Kondo temperature in this regime~\cite{Cronenwett}. 
Additionally, an anomalous dependence of the width of zero-bias anomaly (ZBA) peak on gate voltage is commonly observed below 
$0.7 G_0$~\cite{Cronenwett,Sarkozy,Ren}, further suggesting the specialness of $0.7 G_0$. 
The threshold value of 0.7 may vary slightly depending on the sample.

%%%%%%%%%%%%%%%%%%%%%%%%%%%%%%%%%%%%%%%%%%%%%%%%%%%%%%%%%%%%%%%%%%%%%%%%%%%%%%%%%%%%%%%%
\begin{figure}[b] 
\centering
\includegraphics[width=3.0 in]{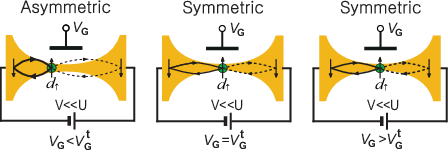}
\caption{Schematic illustration of the entangled state formed by a linear combination of two Kondo singlets, showing the variation in the position of the localized spin and the Kondo coupling strength (indicated by line thickness) as $V_{\rm G}$ increases.
\(V_{\rm G}^{\rm t}\) denotes the threshold at which left--right symmetry begins to break.
Downward arrows on both sides indicate the coherent spins at the Fermi levels of the reservoirs.
}
\label{fig1}
\end{figure}
%%%%%%%%%%%%%%%%%%%%%%%%%%%%%%%%%%%%%%%%%%%%%%%%%%%%%%%%%%%%%%%%%%%%%%%%%%%%%%%%%%%%%%%%

This study aims to address the following fundamental issues: Why does $0.7 G_0$ mark a boundary in QPC behavior? 
What occurs as the side-gate voltage $V_{\rm G}$ crosses the value yielding $G = 0.7 G_0$ in conductance? And finally, do these anomalies signal the presence of a 
quantum phase transition (QPT)?

Recently, QPTs have become an important topic in mesoscopic quantum systems. 
Most research has focused on extended quantum dot structures, such as double quantum dots~\cite{Kleeorin,Kozin,Bargerbos}, or systems showing singlet–triplet transitions in two-stage Kondo regime~\cite{Roch,Guo,Almeida}. 
To the best of our knowledge, however, no QPT has been reported in simpler systems such as a single quantum dot or a QPC.
Unlike a single quantum dot, a QPC—one of the most fundamental quantum transport systems—show a possibility of phase transition in its $V_{\rm G}$-dependent features illustrated in  Fig.~\ref{fig1}.
Therefore, we turn our attention to the QPC and seek to identify a QPT driven by variation in $V_{\rm G}$. 

According to spin-density-functional theory calculations for a QPC~\cite{Rejec,Berggren}, a localized spin forms near the edge of the QPC constriction at low $V_{\rm G}$, and migrates toward the center as $V_{\rm G}$ increases; once it has reached the center, it remains there despite further increments in $V_{\rm G}$.
This evolution implies a spontaneous symmetry breaking in the left–right Kondo coupling strengths as $V_{\rm G}$ crosses a transition point $V_{\rm G}^{\rm t}$, at which the couplings become equal (see Fig.~\ref{fig1}). 
Such symmetry breaking provides a sufficient condition for the occurrence of a continuous phase transition, in which the transition point $V_{\rm G}^{\rm t}$ becomes a critical point $V_{\rm G}^{\rm c}$.

We therefore aim to identify a continuous QPT at $V_{\rm G}^{\rm t}$, which could resolve the longstanding anomalies in QPC behavior.

To this end, we focus on the local density of states (LDOS), which captures the fundamental electronic structure of the system. While obtaining the LDOS in a non-equilibrium steady state is generally challenging, we build on our previous success in replicating the experimentally observed 
$V_{\rm G}$-dependent differential conductance line shapes~\cite{iop-qpc}, which lends feasibility to our approach.

To reveal the QPT, we analyze the zero-temperature quasiparticle energy at the Fermi level, defined as
\begin{equation}
E_{\rm ZBA}(V_{\rm G}) = 2\int_0^{\omega_{\rm min}} \omega \rho_{d\sigma}(\omega,V_{\rm G}) d\omega, 
\label{eq:ZBA energy}
\end{equation}
where $\omega_{\rm min}$ denotes the position of the local minimum in the local density of states (LDOS), $\rho_{d\sigma}(\omega, V_{\rm G})$, characterizing the extent of the ZBA peak.
Here, $\sigma$ represents the spin index, and $d$ refers to the localized energy level.
Equation (\ref{eq:ZBA energy})  offers a consistent measure for comparing the relative magnitude of the quasiparticle energy across different gate voltages.
The gate-voltage dependence of $E_{\rm ZBA}(V_{\rm G})$ reveals hidden aspects of the QPC and serves as a free energy describing QPT. 

The structure of this paper is as follows. Section 2 outlines the theoretical framework used to reproduce the $V_{\rm G}$-dependent differential conductance and discusses the roles of key parameters. Section 3 presents the $V_{\rm G}$-dependent LDOS and quasiparticle energy, along with an analysis confirming the presence of a QPT. 
Finally, section 4 concludes the paper.

\section{Differential conductance and involved parameters} 
Experimentally measured gate-voltage-dependent differential conductance~\cite{Sarkozy}, $\frac{dI}{dV}(V_{\rm G})$ versus $V$, where $I$ is the source-drain current and $V$ the applied bias, was theoretically reproduced in our previous study~\cite{iop-qpc}. 
The differential conductance is given by the expression:
%%%%%%%%%%%%%%%%%%%%%%%%%%%%%%%%%%%%%%%%%%%%%%%%%%%%%%%%%%%%%%%%%%%%%%%%%%%%%%%%%%%%%%%%
\begin{figure}[t] 
\centering
\includegraphics[width=2.4 in]{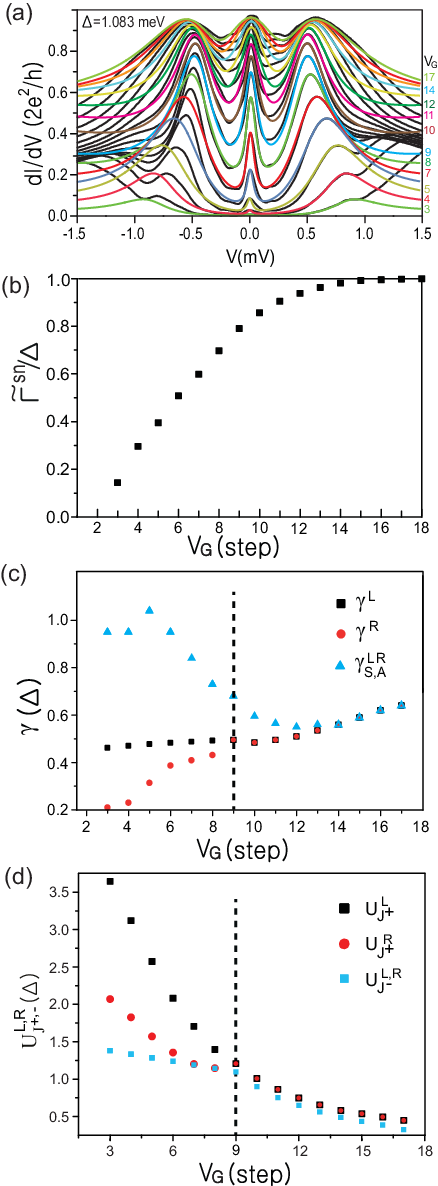}
 \caption{(a) Colored curves represent the theoretical $dI/dV$ line shapes, while black lines correspond to the experimental data~\cite{Sarkozy}. 
$\Delta$ denotes the energy unit.
(b) Behavior of $\widetilde{\Gamma}^{\rm sn}(V_{\rm G})$.
(c) $V_{\rm G}$-dependent behavior of the $\gamma$-parameters.
(d) $V_{\rm G}$-dependent behavior of $U_{J^+}^L$,  $U_{J^+}^R$, and $U_{J^-}^{L,R}$. 
We set $U_{j^-}^L=U_{j^-}^R$ because $j^-$ represents a steady current.
Vertical dashed lines in (c) and (d) mark the point of left--right symmetry breaking. 
}
\label{fig2}
\end{figure}
%%%%%%%%%%%%%%%%%%%%%%%%%%%%%%%%%%%%%%%%%%%%%%%%%%%%%%%%%%%%%%%%%%%%%%%%%%%%%%%%%%%%%%%%

\begin{equation} 
\frac{dI}{dV}(V_{\rm G}) = \left. \frac{2e^2}{h} \widetilde{\Gamma}^{\rm sn}(V_{\rm G}) {\rm Im} {\mathcal{G}}_{dd\uparrow}^{+\rm sn}(\omega, V_{\rm G}) \right|_{\hbar\omega = eV},
\label{Conductance}
\end{equation}  
where $\widetilde{\Gamma}^{\rm sn}(V_{\rm G})$ is the effective coupling function in a two-terminal device (illustrated in Fig.~\ref{fig1}), and 
\({\rm Im} \, {\mathcal{G}}_{dd\uparrow}^{+\rm sn}(\omega, V_{\rm G})\) denotes the imaginary part of the on-site retarded Green’s function at the localized spin site indicated by ${\it d}$. 
The superscript "sn" denotes a steady-state non-equilibrium condition. 
Equation (\ref{Conductance}) serves as a practical adaptation of $V_{\rm G}$-dependent differential conductance derived from the Meir-Wingreen current formula~\cite{Meir}.

In the Liouville space formalism, with which basis operators can be systematically determined~\cite{Hong8}, the retarded Green’s function $\mathcal{G}_{dd\uparrow}^{+\rm sn}(\omega)$ is given by the $dd$ element of the Green's function matrix:
$$
i\mathcal{G}_{dd\uparrow}^{+\rm sn}(\omega) = \left[\frac{1}{z \mathbf{I} + i \mathbf{L}}\right]_{dd},
$$
where $z = -i\omega + 0^+$, $\mathbf{I}$ is the identity operator, and $\mathbf{L}$ is the Liouville operator~\cite{Fulde}. The LDOS at the site of the localized spin is then given by:
$$
\rho_{d\uparrow}(\omega)=-\frac{1}{\pi}{\rm Im}{\mathcal G}^{+}_{dd\uparrow}(\omega)=\frac{1}{\pi}{\rm Re}({\bf M}^{-1})_{dd},
$$
with $\mathbf{M} = z \mathbf{I} + i \mathbf{L}$.

We constructed the matrix $\mathbf{M}$ in terms of the basis operators spanning the working Liouville space and derived an equivalent $5 \times 5$ reduced matrix $\mathbf{M}_r$ as follows:
\begin{eqnarray*}
{\rm\bf M}_r=\left(\begin{array}{c c c c c} -i\omega & \gamma^L &
-U^L_{j^-} & \gamma^{LR}_S & \gamma^{LR}_A \\ -\gamma^L & -i\omega
& -U^L_{j^+} & \gamma^{LR}_A & \gamma^{LR}_S \\
U_{j^-}^{L*} &  U_{j^+}^{L*} & -i\omega &  U^{R*}_{j^+} &
U^{R*}_{j^-} \\  -\gamma^{LR}_S & -\gamma^{LR}_A & -U_{j^+}^R  &
-i\omega & -\gamma^R \\
 -\gamma^{LR}_A &  -\gamma^{LR}_S &  -U_{j^-}^R  & \gamma^R & -i\omega
\end{array}\right)+i{\bf \Sigma}, %\nonumber \\
\end{eqnarray*} 
where the second term is the self-energy matrix. 
This form is obtained by integrating out the reservoir degrees of freedom using a matrix reduction technique~\cite{Lowdin, Mujica}.
Details can be found in  reference~\cite{iop-qpc}.

In this context, the variable $\omega$ is defined as $\omega - \epsilon_d - U \langle n_{d\downarrow} \rangle$, where $\epsilon_d$ is the energy level of the localized state, $U$ the on-site Coulomb interaction, and $\langle n_{d\downarrow} \rangle$ the occupancy probability of the spin-down electron.
The calculated $V_{\rm G}$-dependent differential conductance and the parameters involved are shown in Fig.~\ref{fig2}.

\begin{figure}[t] 
\centering
\includegraphics[width=2.5 in]{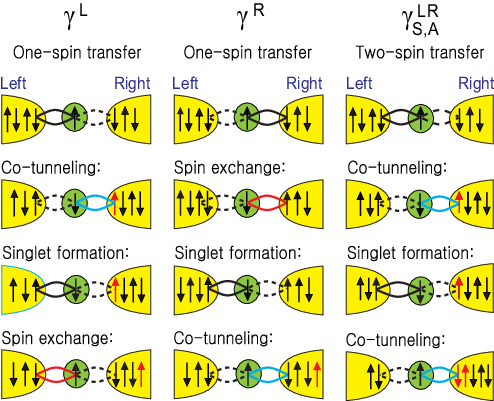}
 \caption{Unidirectional entangled-state tunneling represented by the operator dynamics contained in $\gamma$-parameters.
The unidirectionality gives \(\gamma^{LR}_{S}=\gamma^{LR}_{A}\). }
\label{fig3}
\end{figure}

The key dynamical processes responsible for the coherent current are encoded in the $\gamma$-parameters of $\mathbf{M}_r$. 
These parameters describe cyclic processes of entangled-state tunneling, as illustrated in Fig.~\ref{fig3}. 
Specifically, $\gamma^L$ and $\gamma^R$ involve spin-exchange processes and therefore reflect the Kondo coupling strengths on the left and right singlets, respectively. 
In contrast, $\gamma^{LR}_S$ and $\gamma^{LR}_A$ describe singlet co-tunneling processes that do not involve spin exchange and thus are independent of the Kondo coupling strength.

In our previous work~\cite{iop-qpc}, we identified from atomic limit analysis that $\gamma^L$ and $\gamma^R$ determine the spectral weight of the ZBA peak, 
while $\gamma^{LR}_S$ and $\gamma^{LR}_A$ determine that of the coherent side peaks. 
Additionally, the unidirectional property of entangled-state tunneling yields the equality, \(\gamma^{LR}_{S}=\gamma^{LR}_{A}\).

Meanwhile, the parameters $U_{j^\mp}^{L,R}$ represent effective Coulomb interactions screened by the motion of incoherent spins either incoming from or outgoing to the left and right reservoirs. 
These matrix elements do not capture coherent tunneling phenomena but depend on the side gate voltage $V_{\rm G}$. 

%%%%%%%%%%%%%%%%%%%%%%%%%%%%%%%%%%%%%%%%%%%%%%%%%%%%%%%%%%%%%%%%%%%%%%%%%%%%%%%%%%%%%%%%
%%%%%%%%%%%%%%%%%%%%%%%%%%%%%%%%%%%%%%%%%%%%%%%%%%%%%%%%%%%%%%%%%%%%%%%%%%%%%%%%%%%%%%%%

\section{Quantum phase transition} 
The trends in parameters shown in Figs.~\ref{fig2}(c) and \ref{fig2}(d) clearly indicate a transition between symmetric ($\gamma^L = \gamma^R$; $U^{L}_{j^+}=U^{R}_{j^+}$) and asymmetric ($\gamma^L \neq \gamma^R$; $U^{L}_{j^+} \neq U^{R}_{j^+}$) states at $V_{\rm G} = 9$. 
As illustrated in Fig.~\ref{fig3}, the symmetry and asymmetry in coherent tunneling are reflected in the difference between 
$\gamma^L$ and $\gamma^R$.
This contrast suggests the presence of a QPT as $V_{\rm G}$ crosses the transition point $V_{\rm G} = 9$. 
Identifying solid evidence of such a $V_{\rm G}$-driven QPT in a QPC is both fundamental and compelling.

%%%%%%%%%%%%%%%%%%%%%%%%%%%%%%%%%%%%%%%%%%%%%%%%%%%%%%%%%%%%%%%%%%%%%%%%%%%%%%%%%%%%%%%%
\begin{figure}[t] 
\centering
\includegraphics[width=2.5 in]{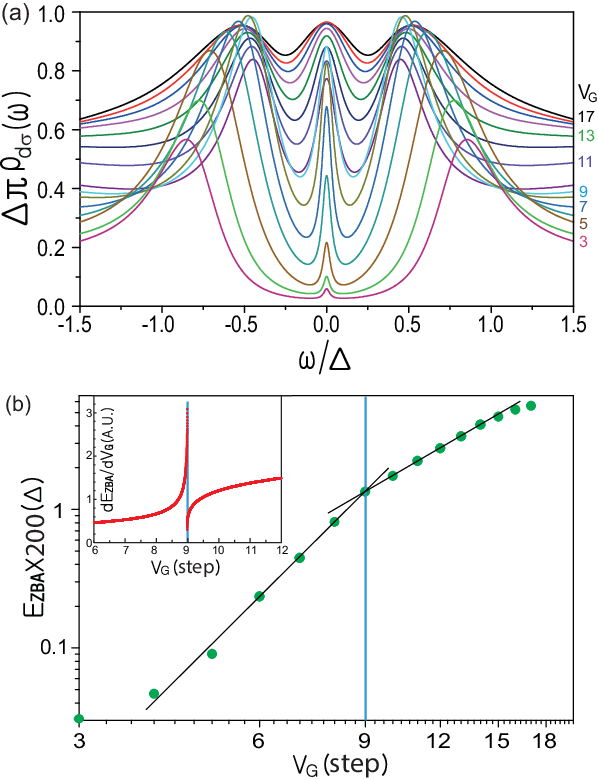}
 \caption{(a) $V_{\rm G}$-dependent LDOS.  
(b) Behaviors of the ZBA energy plotted on a logarithmic scale.
Inset: Derivative of the ZBA energy as a function of $V_{\rm G}$ plotted on a linear scale.}
\label{fig4}
\end{figure}
%%%%%%%%%%%%%%%%%%%%%%%%%%%%%%%%%%%%%%%%%%%%%%%%%%%%%%%%%%%%%%%%%%%%%%%%%%%%%%%%%%%%%%%%

To this end, we obtain the $V_{\rm G}$-dependent LDOS by dividing the differential conductance $dI/dV$, expressed in units of $2e^2/h$ as shown in 
Fig.~\ref{fig2}(a), by the effective coupling $\widetilde{\Gamma}^{\rm sn}(V_{\rm G})$ presented in Fig.~\ref{fig2}(b). 
The resulting LDOS is shown in Fig.~\ref{fig4}(a).

Next, we compute the $V_{\rm G}$-dependent quasiparticle energy at the Fermi level at zero temperature using equation~(\ref{eq:ZBA energy}). 
The results, plotted on a logarithmic scale in Fig.~\ref{fig4}(b), align remarkably well with two distinct linear regimes, indicating separate power-law behaviors:
$$ E_{\rm ZBA}/\Delta = 0.00671 - 0.00266 \times |V_{\rm G} - 9|^{0.666}, \quad (V_{\rm G} \leq 9) $$
$$ E_{\rm ZBA}/\Delta = 0.00671 + 0.00200 \times |V_{\rm G} - 9|^{1.169}, \quad (V_{\rm G} \geq 9), $$
which serve as compelling evidence of a QPT. 
Furthermore, the derivative $\partial E_{\rm ZBA}/\partial V_{\rm G}$, shown in the inset of 
Fig.~\ref{fig4}(b), diverges as $V_{\rm G} \to V_{\rm G}^{\rm c-}$ and vanishes as $V_{\rm G} \to V_{\rm G}^{\rm c+}$.
These results confirm a continuous phase transition. 

Notably, the critical point corresponds to a linear conductance of $0.7 G_0$, as inferred from the ZBA peak maximum in Fig.~\ref{fig2}(a) at $V_{\rm G}=9$. 
Furthermore, the pronounced difference at the transition point, shown in the inset of Fig.~\ref{fig4}(b), suggests the potential for developing quantum sensors.

%%%%%%%%%%%%%%%%%%%%%%%%%%%%%%%%%%%%%%%%%%%%%%%%%%%%%%%%%%%%%%%%%%%%%%%%%%%%%%%%%%%%%%%
\begin{figure}[t] 
\centering
\includegraphics[width=3.0 in]{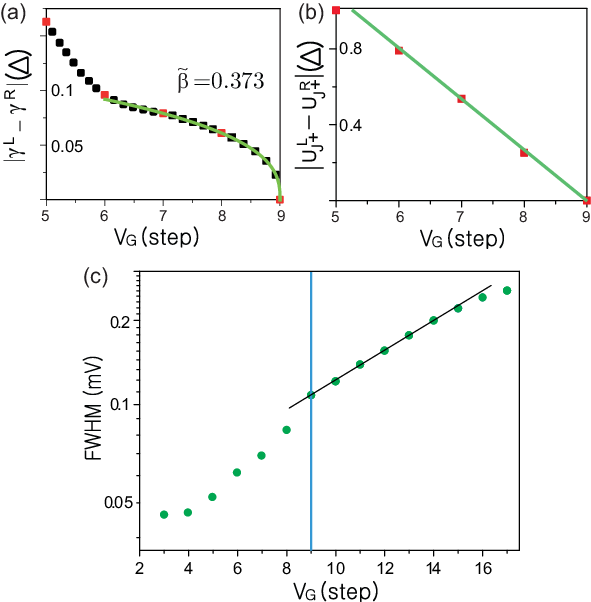}
 \caption{(a) Critical behavior of the order parameter. 
Red squares represent the data, while dark squares are added as visual guides.
(b) Behavior of \(|U^{L}_{j^+}-U^{R}_{j^+}|\) as a function of  $V_{\rm G}$. (c) ZBA width plotted on a semilogarithmic scale. }
\label{fig5}
\end{figure}
%%%%%%%%%%%%%%%%%%%%%%%%%%%%%%%%%%%%%%%%%%%%%%%%%%%%%%%%%%%%%%%%%%%%%%%%%%%%%%%%%%%%%%%%

In the context of continuous phase transitions, identifying an appropriate order parameter is crucial. 
As established in our previous work~\cite{iop-qpc}, the parameters $\gamma^L$ and $\gamma^R$ are distinguished not by differences in hybridization strength, but by the spatial locations at which spin exchange occurs, as illustrated in Fig.~\ref{fig3}.
Accordingly, the quantity $|\gamma^L - \gamma^R|$ emerges as a natural order parameter, capturing the asymmetry in Kondo coupling strength between the left and right reservoirs and the localized spin situated between them.
The critical behavior of this order parameter is expressed by the relation:
$$ |\gamma^L - \gamma^R| \propto |V_{\rm G} - V_{\rm G}^{\rm c}|^{\tilde\beta}, $$
with a critical exponent $\tilde{\beta} = 0.373$, as shown in Fig.~\ref{fig5}(a). 
For comparison, Fig.~\ref{fig5}(b) displays $|U^{L}_{j^+} - U^{R}_{j^+}|$ against $V_{\rm G}$, illustrating that this quantity does not follow the critical behavior of order parameter.

Lastly, we discuss the behavior of the full width at half maximum (FWHM) of the ZBA peak.
It is widely known that the FWHM of the Kondo resonance peak amounts to twice of the Kondo temperature.
Therefore, the discussion of the FWHM of the ZBA provides information on the behavior of the Kondo temperature.

For a conventional single-impurity Kondo system described by the single-reservoir asymmetric Anderson model, scaling theory predicts that the Kondo temperature, $T_{\rm K}$, obeys the relation $T_{\rm K}\propto {\rm exp}[\pi\varepsilon_0(\varepsilon_0+U)/\Gamma U]$, where $\varepsilon_0$ denotes the energy of the localized state, $\varepsilon_0+U$ the energy of the excited state, and $\Gamma$, the width of the Coulomb peak of strength $U$~\cite{Haldane}.
This relation underpins the experimentally observed behavior of the Kondo temperature: $\ln(T_{\rm K}) \propto |V_{\rm G} - V_{\rm G}^0|$~\cite{Cronenwett,Chen}.
However, this linear dependence holds only over a limited range of gate voltages $V_{\rm G} < V_{\rm G}^0$ (or equivalently $G < 0.7 G_0$), which has long been considered a puzzling feature in QPCs.

Since the scaling analysis presumes a single Kondo temperature for the underlying system, the coexistence of two Kondo temperatures introduces an ambiguity, leading to anomalous behavior. 
This is clearly reflected in the FWHM data plotted on a semilogarithmic scale in Fig.~\ref{fig5}(c).
A clear linear trend is observed only in the symmetric regime ($V_{\rm G} \geq 9$), where a single Kondo temperature is present.
In contrast, data in the asymmetric regime ($V_{\rm G} < 9$) deviate from this trend, signaling a breakdown of conventional scaling.

These results delineate the valid regime of scaling analysis as $V_{\rm G} \geq V_{\rm G}^c$, or equivalently, $G \geq 0.7 G_0$.
The irregular features observed on the left side of Fig.~\ref{fig5}(c) are consistent with earlier experimental findings for $G < 0.7 G_0$, reported in references~\cite{Cronenwett, Sarkozy, Ren}.
Thus, the longstanding “scaling puzzle” in QPCs is resolved by recognizing the emergence of two distinct Kondo temperatures in the asymmetric phase.
\section{Conclusion} 
We have identified a gate-voltage-driven continuous quantum phase transition occurring at the critical value $V_{\rm G}^{\rm c} = 9$, corresponding to a conductance of $0.7 G_0$, based on the $V_{\rm G}$-dependent differential conductance obtained in our previous study~\cite{iop-qpc}. 
This transition delineates two distinct phases: a symmetric phase for $V_{\rm G} \geq V_{\rm G}^{\rm c}$ (or $G \geq 0.7 G_0$), characterized by left–right symmetric Kondo couplings and a single Kondo temperature; and an asymmetric phase for $V_{\rm G} < V_{\rm G}^{\rm c}$ (or $G < 0.7 G_0$), in which left–right asymmetric Kondo couplings give rise to two distinct Kondo temperatures.

This finding resolves several longstanding puzzles in quantum point contacts, including the indeterminate Kondo temperature, the deviation of scaled conductance data from the scaling function~\cite{Cronenwett}, and the anomalous gate-voltage dependence of the width of the zero-bias anomaly~\cite{Cronenwett,Sarkozy,Ren}. 
Notably, these anomalous behaviors are confined to the low-conductance regime ($G < 0.7 G_0$), now understood as the asymmetric Kondo phase.

\vspace{0.5 cm}

\end{document}